\documentclass[english,two column]{revtex4-1}
\usepackage[LGR,T1]{fontenc}
\usepackage[latin9]{inputenc}
\usepackage{amsmath}
\usepackage{amssymb}
\usepackage{graphicx}
\usepackage{hyperref}
\hypersetup{breaklinks,
colorlinks=true,%
citecolor=blue,%
filecolor=black,%
linkcolor=blue,%
urlcolor=blue}

\makeatletter

\ProvideTextCommand{\~}{LGR}[1]{\char126#1}

\makeatother

\usepackage{babel}
\begin{document}
\title{Novel dynamical excitations and roton-based measurement of Cooper-pair momentum in a two-dimensional Fulde-Ferrell-Larkin-Ovchinnikov superfluid on optical lattices}
\author{Shuning Tan$^{1}$}
\author{Jiayi Shi$^{1}$}
\author{Peng Zou$^{2}$}
\email{phy.zoupeng@gmail.com}
\author{Tianxing Ma$^{3}$}
\email{txma@bnu.edu.cn}
\author{Huaisong Zhao$^{2}$}
\email{hszhao@qdu.edu.cn}
\affiliation{$^{1}$Key Laboratory for Microstructural Material Physics of Hebei Province, School of Science, Yanshan University, Qinhuangdao 066004, P. R. China.}
\affiliation{$^{2}$Centre for Theoretical and Computational Physics, College of Physics, Qingdao University, Qingdao 266071, P. R. China}
\affiliation{$^{3}$School of Physics and Astronomy, Beijing Normal University, Beijing 100875, P. R. China}
\begin{abstract}
Determining the center-of-mass (COM) momentum of Cooper pairs in unconventional superconductors or superfluids is a topic of great interest in condensed matter physics and ultracold atomic gases. Theoretically, we investigate the dynamical excitations of a two-dimensional spin-polarized attractive Hubbard model on a square
optical lattice under an effective Zeeman field by computing the density and spin dynamical structure factors, focusing on phase transition from a Bardeen-Cooper-Schrieffer (BCS) superfluid to an Fulde-Ferrell-Larkin-Ovchinnikov (FFLO) superfluid. In the FFLO superfluid, besides the phonon mode in the density channel, a low-energy bogolon mode emerges in the spin channel, which is associated with Bogoliubov quasiparticles on a Bogoliubov Fermi
surface. Moreover, the dynamical excitations exhibit pronounced anisotropy in momentum space due to the finite COM momentum.
At half filling, the roton mode around $[\pi,\pi]$ evolves from a point-like minimum into a ring structure shifted by the COM momentum across the BCS-FFLO transition, providing a roton-based protocol to extract the COM momentum. These predictions provide key insights for confirming the existence of FFLO superfluids and understanding their dynamical excitation spectra.
\end{abstract}
\maketitle

\section{Introduction}
  Unlike conventional Bardeen-Cooper-Schrieffer (BCS) superconductors, where Cooper pairs carry zero center-of-mass (COM) momentum ${\bf Q}=0$, there exists a class of exotic superconductors with finite COM momentum, such as the pair-density wave (PDW) superconductors and Fulde-Ferrell-Larkin-Ovchinnikov (FFLO) superconductors.  The PDW state arises without an applied Zeeman field while the Fulde-Ferrell-Larkin-Ovchinnikov (FFLO) state is stabilized under an applied Zeeman field.
 Since the experimental realization of the PDW in various superconductors \cite{Agterberg2020,Chen2021,Liu2023,Zhaoh2023,Kong2025,Zhang2024,Kittaka2023,Kasahara2020,Han2025}, the physical properties of PDW system have attracted significant attention \cite{Panigrahi2025,Lamponen2025}. Meanwhile, the FFLO state have been discussed in the multiband iron-based superconductors \cite{Kasahara2020,Kasahara2021,Zhou2021,Liu2025}.  As highly tunable and clean systems, the ultracold atomic gases provide an excellent platform for simulating such novel many-body physics in condensed matter physics \cite{Bloch2008rmp,Dalibard2011}. In particular, the FFLO superfluid state are widely expected to exist in polarized Fermi gases \cite{Kinnunen2018,Radzihovsky2010,Sheehy2007,Mora2005,Batrouni2008,Liuxjj2013,Dong2013,Koponen2007,Huh2006,Xub2014}, motivating extensive experimental and theoretical efforts to realize it in ultracold atomic gases \cite{Liu2007,Huhui2018,Zhang2013,xu2014,Liuxj2013,Cao2014,Kawamura2022,Pini2021,Wei2018}. Understanding how the characteristic physical quantities evolve across the BCS-FFLO transition becomes a key issue. Our previous studies exhibited some significant differences on dynamical excitations between a conventional BCS superfluid and various exotic superfluid phases, including topological superfluid \cite{Zhao2023,Gao2023,Zhao2024}, Sarma phase \cite{Zou2021}, and one-dimensional FFLO gases \cite{Zoup2024}.
  Recent progress shows that the collective mode can distinguish PDW and charge-density-wave (CDW) states \cite{WuY2025}. These studies indicate that dynamical excitations can serve as a sensitive tool for identifying the FFLO superfluids and discriminating them from the conventional BCS superfluid.

In conventional BCS superfluids, the Fermi surface disappears due to the opening of the pairing gap, so that the single-particle excitations are gapped and appear only above a finite threshold energy. Within this energy range, the single-particle excitations and collective modes can be clearly separated. As a result, the collective phonon mode can be measured distinctly. However, in the FFLO superfluids, the Fermi surface persists, and the quasiparticle spectrum remain gapless and form a  Bogoliubov Fermi surface (BG-FS) \cite{Autti2020,Jiang2021,Agterberg2017,Miki2021,Brydon2018,Hoshino2022}, where the low-energy quasiparticle excitations near the Fermi energy are described by the Bogoliubov quasiparticles (bogolons). This Bogoliubov Fermi liquid exhibits distinct physical properties, such as the odd-frequency pairing \cite{Miki2021,Bergeret2005,Linde2019}, the unconventional impurity effect on the single-particle spectra and density of states (DOSs)\cite{Hoshino2022}, a Drude-like optical conductivity contrasting with that of d-wave superconductors \cite{oh2021}.Therefore, in addition to the conventional single-particle excitations, the low-energy dynamical excitations are composed of the phonon and bogolon. Moreover, in the FFLO superfluids, the single-particle excitations strongly compete with these collective modes. Hence, this competition makes it essential to simultaneously calculate the full dynamical excitations. These dynamical excitations can be probed through the dynamical structure factors, which are two-body correlation observables \cite{Combescot2006,Watabe2010,Zhao2020}. In the ultracold atomic gases, the dynamical structure factors can be directly measured using two-photon Bragg spectroscopy \cite{Pagano2014,Sobirey2022,Veeravalli08,Hoinka17,Biss2022,Senaratne2022,Li2022,Dyke2023}, and can also be simulated through various numerical approaches \cite{Vitali2020,Vitali2022,Apostoli2024,Lohofer2015}.

Optical lattices have become a widely used platform for simulating the physical models in condensed matter physics, such as the Fermi-Hubbard model \cite{Jrdens2008,Schneider2008,Greif13,Parsons16,Cheuk16,Koepsell2021,Boll16,Brown17,Wu2016,Arovas2022,Hartke2023}.
The attractive Fermi-Hubbard model is strongly associated with the superfluid phenomenon and has been extensively studied through numerous experiments \cite{Mitra2018,Peter2020,Hackermuller2010,Gall2020,Schneider2012} and theoretical studies \cite{Mondaini2015,Cocchi2016,Strohmaier07,Moreo07,Gukelberger16,Shenoy2008}. Compared with the continuous gases characterized by an extremely narrow parameter window, the FFLO states on an optical lattice exhibit significantly broader parameter tunability \cite{Koponen2007}, thereby enhancing the experimental feasibility. The collective modes of the lattice FFLO superfluid have been investigated through several numerical simulations \cite{Edge2009,Edge2010,Heikkinen2011,Koinov2011,Huang2022}, revealing that the collective modes exhibits the anisotropic behavior. For instance, the sound speed parallel to ${\bf Q}$ is greater than that perpendicular to ${\bf Q}$ \cite{Heikkinen2011}, and two asymmetric roton-like collective modes have been identified \cite{Koinov2011}. Nevertheless, until now, a systematic investigation of the full dynamical excitations is still lacking, particularly concerning the phonon, bogolon, roton modes, as well as their competition with the single-particle excitation continua. In particular, an experimental protocol to extract the COM momentum ${\bf Q}$ from the dynamical excitations remains an open question.

 In this paper, we theoretically compute the density and spin dynamical structure factors of a two-dimensional (2D) spin-polarized attractive Fermi-Hubbard model on a square lattice and investigate the evolution of the full dynamical excitations across the BCS-FFLO transition within the random phase approximation (RPA) \cite{Anderson1958,Liu2004,He2016,Ganesh2009}. A central result is a roton-based measurement protocol for the Cooper-pair momentum. At half filling, the roton mode near $[\pi,\pi]$ in density dynamical structure factor enables a direct extraction of the COM momentum magnitude from the roton minimum.
In ultracold atomic gases, the response theories based on RPA have been demonstrated to describe the dynamical excitations reliably. For three-dimensional(3D) Fermi superfluids, they even yield quantitatively consistent results when compared with a two-photon Bragg scattering spectra \cite{Zou2018,Zou2010,Biss2022}, and in 2D system they reproduce qualitative results found in quantum Monte Carlo calculations \cite{Zhao2020,Vitali2020,Zhao2023-3}. We therefore expect that our theoretical strategy provides a qualitatively accurate prediction for a 2D lattice system in the intermediate-coupling regimes \cite{Qin2022}.

This paper is organized as follows. In Sec \ref{modelH}, we derive mean-field Green's functions of 2D spin-polarized attractive Fermi-Hubbard model on a square lattice using the equations of motion approach. In Sec. \ref{RPADSF}, we formulate the RPA response functions and obtain the dynamical structure factors.  The results of the dynamical structure factors are presented and discussed in Sec. \ref{DSFHalffiling}, focusing on the anisotropic collective modes and complex single-excitations at half-filling. In Sec. \ref{DSFDoping}, we analyze the doping dependence of the dynamical excitations. A discussion is given in Sec. \ref{discussion}, and the conclusions are summarized in Sec. \ref{summary}. Additional calculation details are provided in the Appendix.

\section{Model and Hamiltonian}
\label{modelH}
For a polarized two-component 2D attractive Fermi-Hubbard model on an optical lattice, an FFLO superfluid state may exist under proper interaction strength and external Zeeman field. Consequently, this system becomes an effective platform to investigate the physical properties of the FFLO superfluids. Its Hamiltonian in momentum space is given by:
\begin{equation}
H = \sum_{{\bf k},\sigma}\xi_{{\bf k}\sigma}
       C_{{\bf k}\sigma}^\dagger C_{{\bf k}\sigma}
    - U\sum_{{\bf k}}
       C_{{\bf k}\uparrow}^\dagger
      C_{{\bf Q-k}\downarrow}^\dagger
      C_{{\bf Q-k}\downarrow}
      C_{{\bf k}\uparrow},
\label{eq:H0}
\end{equation}
where $\xi_{{\bf k}\sigma}=\varepsilon_{\bf k}-h\sigma_{z}$, $\varepsilon_{\bf k}=-Zt\gamma_{\bf k}-\mu$, $\gamma_{\bf k}=\left(\cos{k_{x}}+\cos{k_{y}}\right)/2$, and $\sigma_{z}=1(-1)$ for spin $\sigma=\uparrow (\downarrow)$. Here, $\mu$ is the average chemical potential and $h$ is the effective Zeeman field. For a 2D square lattice, the coordination number is $Z=4$. $C_{{\bf k}\sigma}$ and $C_{{\bf k}\sigma}^\dagger$ denote the annihilation and creation operators for a fermion with momentum ${\bf k}$ and spin \(\sigma = \uparrow, \downarrow\). The vector ${\bf Q}$ denotes the COM momentum. The on-site attractive interaction between the opposite-spin atoms is characterized by the Hubbard energy $U>0$. In the following discussions, we take $U$ as the unit of energy and the lattice constant $a_{0}$ as the unit of length.

In an FF superfluid state, the pairing gap (the order parameter) is defined as, $\Delta^{*}=U<C_{{\bf k}\uparrow}^{\dagger}C_{{\bf Q-k}\downarrow}^{\dagger}>$, and take the plane wave form, $\Delta({\bf r})=\Delta e^{i{\bf Q \cdot r}}$.
 Within the mean-field approximation, the four-operator interaction term is decoupled into a two-operator form:
$UC_{{\bf k}\uparrow}^{\dagger}C_{{\bf Q-k}\downarrow}^{\dagger}C_{{\bf Q-k}\downarrow}C_{{\bf k}\uparrow}=\Delta^{*}C_{{\bf Q-k}\downarrow}C_{{\bf k}\uparrow}+\Delta C_{{\bf k}\uparrow}^{\dagger}C_{{\bf Q-k}\downarrow}^{\dagger}-\Delta^{2}/U$.
 Consequently, the mean-field Hamiltonian is obtained as, $H_{MF}= \sum_{{\bf k},\sigma} \xi_{{\bf k}\sigma} C_{{\bf k}\sigma}^{\dagger}C_{{\bf k}\sigma}+\frac{\Delta^2}{U}-\sum_{{\bf k}} (\Delta C_{{\bf k}\uparrow}^{\dagger}C_{{\bf Q-k}\downarrow}^{\dagger}+\Delta^{*} C_{{\bf Q-k}\downarrow}C_{{\bf k}\uparrow})$. We define the spin-up (spin-down) particle number density Green's function, $G_{\uparrow}({\bf k},\tau-\tau{'})=-\left\langle T_{\tau} C_{{\bf k}\uparrow}(\tau)C^{\dagger}_{{\bf k}\uparrow}(\tau{'})\right\rangle$ ($G_{\downarrow}({\bf k},\tau-\tau{'})=-\left\langle T_{\tau} C_{{\bf k}\downarrow}(\tau)C^{\dagger}_{{\bf k}\downarrow}(\tau{'})\right\rangle$), and the singlet pairing one $\Gamma^{\dagger}({\bf k},\tau-\tau{'})=-\left\langle T_{\tau} C^{\dagger}_{{\bf Q-k}\downarrow}(\tau)C^{\dagger}_{{\bf k}\uparrow}(\tau{'})\right\rangle$, respectively. Due to the Zeeman field, $G_{\uparrow}\neq G_{\downarrow}$. Using the equations of motion for these Green's functions, these three Green's functions are solved as:
\begin{subequations}\label{bcsgreen}
 \begin{eqnarray}
G_{\uparrow}\left({\bf k},\omega\right)&=&\frac{U_{\bf k}^2}{\omega-E^{(1)}_{\bf k}}+\frac{V_{\bf k}^2}{\omega+E^{(2)}_{\bf k}}\\
G_{\downarrow}\left({\bf k},\omega\right)&=&\frac{U_{\bf k}^2}{\omega-E^{(2)}_{\bf Q-k}}+\frac{V_{\bf k}^2}{\omega+E^{(1)}_{\bf Q-k}}\\
\Gamma^{\dagger}\left(\bf{k},\omega\right)&=&\frac{\Delta^{*}}{2E_{\bf{k}}}
\left(\frac{1}{\omega-E^{(1)}_{\bf k}}-\frac{1}{\omega+E^{(2)}_{\bf k}}\right),
\end{eqnarray}
\end{subequations}
where $U_{\bf k}^2=0.5[1+0.5(\xi_{\bf k}+\xi_{\bf Q-k})/E_{\bf{k}}]$, $V_{\bf k}^2=0.5[1-0.5(\xi_{\bf k}+\xi_{\bf Q-k})/E_{\bf{k}}]$. The quasiparticle spectra $E^{(1)}_{\bf k}=E_{\bf{k}}+(\xi_{\bf k}-\xi_{\bf Q-k})/{2}-h$, $E^{(2)}_{\bf k}=E_{\bf{k}}-(\xi_{\bf k}-\xi_{\bf Q-k})/{2}+h$, where $E_{\bf{k}}=\sqrt{{(\xi_{\bf k}+\xi_{\bf Q-k})}^2/4+{\Delta^{2}}}$. Furthermore, the dispersion of $-E^{(2)}_{\bf k}$ always lies below the Fermi energy, while $E^{(1)}_{\bf k}$ crosses the Fermi energy.

 Based on the mean-field Hamiltonian, the partition function of a system is given by $Z={\rm T_{r}}[e^{-\beta{H}}]$, where $\beta=1/T$ is the inverse temperature, ${\rm T_{r}}$ denotes the trace.  Thus, the mean-field thermodynamic potential $\Omega=-\ln{Z}/\beta$ in an FFLO state is given by,
\begin{eqnarray}\label{thermalpotential}
\Omega&=&\sum_{\bf k}\left(\frac{\xi_{\bf k}+\xi_{\bf Q-k}}{2}-E_{\bf k}\right)-\frac{\Delta^2}{U}\nonumber\\
&-&T\sum_{\bf k}\ln{(1+e^{-\beta E^{(1)}_{\bf k}})(1+e^{-\beta E^{(2)}_{\bf k}})}.
\end{eqnarray}
When $h$ exceeds the critical value $h_{\rm c}$, the minimum of the thermodynamic potential shifts from $Q=0$ to a finite $Q>0$, signaling a phase transition from a conventional BCS superfluid to an FFLO superfluid. The parameters $\mu$, $\Delta$, and $Q$ are consequently solved by using the stationary conditions, namely, $N=-\partial{\Omega}/\partial{\mu}$, $\partial{\Omega}/\partial{\Delta}=0$, $\partial{\Omega}/\partial{Q}=0$, respectively. These three self-consistent equations can be obtained by
\begin{subequations}\label{threeequtions}
\begin{eqnarray}
N&=&\sum_{\bf k}\left[1-D({\bf k},T)\frac{\xi_{\bf k}+\xi_{\bf Q-k}}{2E_{\bf k}}\right],\\
-\frac{1}{U} &=& \sum_{\bf k}\frac{D({\bf k},T)}{2E_{\bf k}},\\
0&=&\sum_{\bf k}D({\bf k},T)(1-\frac{\xi_{\bf k}+\xi_{\bf Q-k}}{2E_{\bf k}})\frac{\partial{\xi_{\bf Q-k}}}{\partial{Q}},
\end{eqnarray}
\end{subequations}
where the function $D({\bf k},T)=1-n_{F}(E^{(1)}_{\bf k})-n_{F}(E^{(2)}_{\bf k})$ with the Fermi distribution function $n_{F}(x)=1/(e^{\beta{x}}+1)$.
To avoid numerical divergence caused by zeros of $E^{(1)}_{\bf k}$, We set temperature to a typical low value $T=0.01{\varepsilon_{F}}$, approaching the zero-temperature limit. We use a {\it plane polar coordinate system} with $Q$ aligned along the polar axis (${\bf Q}=[Q,0]$) and $\varphi$ as the polar angle. Moreover, $\partial{\xi_{\bf Q-k}}/\partial{Q}=2t\sin(Q-k\cos{\varphi})$, where $\varphi$ denotes the angle between ${\bf k}$ and ${\bf Q}$ and is integrated in the summation.

\begin{figure}[h!]
\includegraphics[scale=0.45]{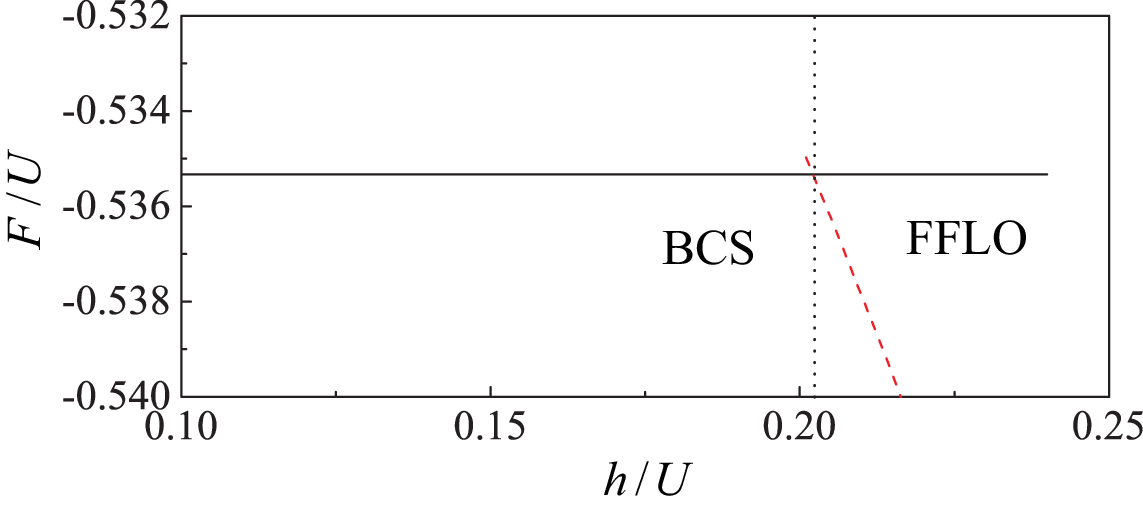}
\caption{Free energy $F=\Omega+{\mu}N$ as a function of $h$ for $t/U=0.3$, respectively. The vertical dotted line marks the BCS-FFLO superfluid phase transition at $h_{\rm c}/U=0.202$.
\label{fig1}}
\end{figure}

The ground state of the system is determined by minimizing the free energy, $F=\Omega+{\mu}N$. In Fig. \ref{fig1}, We plot $F$ as a function of $h$.
Our calculations demonstrate that the ground state corresponds to a conventional BCS superfluid with $Q=0$ at low $h$.  When $h>h_{\rm c}$, the FFLO superfluid with finite $Q$ has a lower free energy than the BCS superfluid, indicating a first-order phase transition. This make the FFLO superfluid become a more stable phase, in which pairing coexists with a finite magnetization. Both the chemical potential and the pairing gap change abruptly at $h_{\rm c}$. Near $h_{\rm c}/U=0.202$, the pairing gap drops from $\Delta_{\rm BCS}/U=0.29$ in the BCS state to $\Delta_{\rm FFLO}/U=0.097$ in the FFLO side, where the corresponding COM momentum is $Q = 0.82$.

\section{Dynamical structure factor within the random phase approximation}
\label{RPADSF}
The mean-field approximation neglects the quantum fluctuations. To incorporate the quantum fluctuations, the random phase approximation (RPA) provides a reliable framework for calculating response functions beyond the mean-field level \cite{Liu2004,He2016,Ganesh2009}.
We briefly outline the main idea of RPA theory for investigating dynamical excitations.  In an FFLO superfluid, four distinct density operators are relevant: the normal spin-up/down densities $\hat{n}_{1}=C^{\dagger}_{\uparrow}C_{\uparrow}$ and $\hat{n}_{2}=C^{\dagger}_{\downarrow}C_{\downarrow}$, together with the anomalous pairing operator and its complex conjugate $\hat{n}_{3}=C_{\downarrow}C_{\uparrow}$, $\hat{n}_{4}=C^{\dagger}_{\uparrow}C^{\dagger}_{\downarrow}$ that describe the Cooper pairing. These four densities are coupled by atomic interactions. A perturbations in any one of them induces fluctuations in the others. Within the linear response theory, a weak external perturbation potential $V_{\rm ext}$ perturbs the system and induces density fluctuations $\delta n$. The corresponding response function is defined as:
\begin{eqnarray}\label{vext}
\delta n=\chi V_{\rm ext},
\end{eqnarray}
where $\delta{\rm n}=[\delta{\rm n}_{1}, \ \delta{\rm n}_{2}, \ \delta{\rm n_{3}}, \ \delta{\rm n_{4}}]^{T}$ and $V_{\rm ext}=[V_{1}, \ V_{2}, \ V_{3}, \ V_{4}]^{T}$. Within the RPA framework, an effective potential is defined as  $V_{\rm eff} \equiv V_{\rm ext}+\delta V^{\rm SC}$. Here, $\delta V^{\rm SC}$ is the self-generated mean-field potential $\delta V^{\rm SC}$ \cite{Liu2004} induced by the density fluctuations, given by
\begin{eqnarray}\label{deVsc}
\delta V^{\rm SC}=U\int d^{2}{\bf r}\left[\delta {\rm n_{4}}\hat{n}_3+\delta {\rm n_{3}}\hat{n}_4
 + \delta {\rm n}_{1}\hat{n}_2+\delta {\rm n}_{2}\hat{n}_1\right].
\end{eqnarray}
The density fluctuation $\delta n$ is related to the effective potential $V_{\rm eff}$ by
\begin{eqnarray}\label{veff}
\delta n=\chi^{0} V_{\rm eff},
\end{eqnarray}
where $\chi^0$ denotes the mean-field response function matrix, which can be readily computed.
Beyond mean-field level, the response function $\chi$ within the RPA approach is connected to its mean-field counterpart $\chi^0$ by
 \begin{eqnarray}\label{chi}
 \chi({\bf q},i\omega_{n})=\frac{\chi^{0}({\bf q},i\omega_{n})}{\hat{1}-\chi^{0}({\bf q},i\omega_{n})U{M_{I}}}.
\end{eqnarray}
Here $M_{I}=\sigma_{0}\otimes\sigma_{x}$ is the direct product of unit matrix $\sigma_{0}$ and the Pauli matrix $\sigma_{x}$. The mean-field response function $\chi^0$ is easy to obtain, and its expression is a $4\times4$ matrix as,
\begin{eqnarray}\label{matrix}
\chi^{0}({\bf q},i\omega_{n})=\left[
\begin{array}{cccccc}
&\chi^{0}_{11}&\chi^{0}_{12}&\chi^{0}_{13}&\chi^{0}_{14}\\
&\chi^{0}_{21} &\chi^{0}_{22}&\chi^{0}_{23}&\chi^{0}_{24}\\
&\chi^{0}_{31}&\chi^{0}_{32}&\chi^{0}_{33}&\chi^{0}_{34}\\
&\chi^{0}_{41}&\chi^{0}_{42}&\chi^{0}_{43}&\chi^{0}_{44}\\
\end{array}
\right].
 \end{eqnarray}
 The dimension of $\chi^0$ reflects the coupling among the four density channels. These 16 matrix elements are given by the density-density correlation functions derived from the defined Green's functions defined earlier. For example, $\chi^{0}_{12}=-\left\langle T_{\tau} \psi^{\dagger}_{\uparrow}({\bf r},\tau)\psi_{\uparrow}({\bf r},\tau)\psi^{\dagger}_{\downarrow}({\bf r'},\tau{'})\psi_{\downarrow}({\bf r'},\tau{'})\right\rangle$, which, after contraction via Wick's theorem, reduces to $\chi^{0}_{12}=-\Gamma^{\dagger}({\bf r}-{\bf r'},\tau-\tau')\Gamma({\bf r'}-{\bf r},\tau'-\tau)$.
 Owing to the symmetries of the system, only 9 matrix elements are independent, i.e.,
 $\chi^{0}_{12}=\chi^{0}_{21}=-\chi^{0}_{33}=-\chi^{0}_{44}$,
 $\chi^{0}_{31}=\chi^{0}_{14}$, $\chi^{0}_{32}=\chi^{0}_{24}$
 $\chi^{0}_{42}=\chi^{0}_{23}$, $\chi^{0}_{41}=\chi^{0}_{13}$.
 Their explicit analytical forms are provided in the Appendix.

The total density (spin) response function $\chi_D$ ($\chi_S$) is defined as $\chi_D \equiv\chi_{11}+\chi_{12}+\chi_{21}+\chi_{22}$ ($\chi_S\equiv\chi_{11}-\chi_{12}-\chi_{21}+\chi_{22}$).
 According to the fluctuation-dissipation theory, the density (spin) dynamical structure factor $S_{D}({\bf q},{\omega})$ ($S_{S}({\bf q},{\omega})$) is given by,
\begin{eqnarray}\label{sqw}
  S_{D/S}({\bf q},{\omega})&=&-\frac{1}{\pi}\frac{1}{1-e^{-\omega/T}}{\rm Im}\chi_{D/S}\left({\bf q},i\omega_{n}\to \omega+i\delta\right),\nonumber\\
 \end{eqnarray}
where ${\bf q}$ and $\omega$ denote the transferred momentum and energy, respectively. The parameter $\delta$ is a small positive number introduced in numerical calculations, typically set to $\delta=0.003$.

\section{Results at half-filling}
 \label{DSFHalffiling}
\begin{figure}[h!]
\centering
\includegraphics[width=0.48\textwidth]{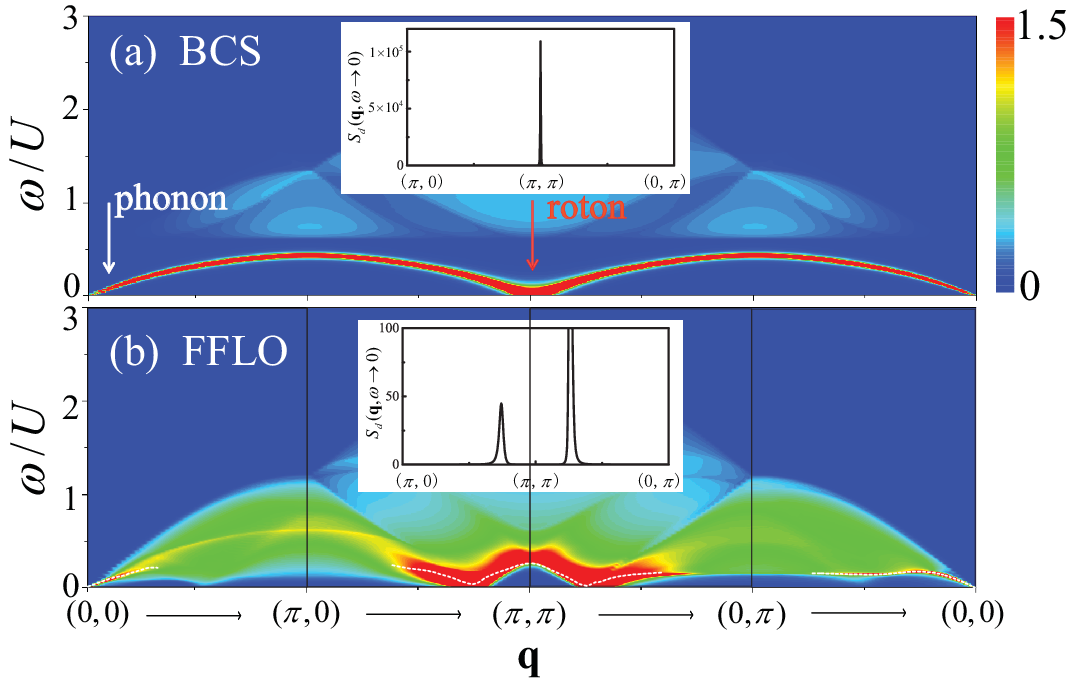}
\caption{\label{DDSF} Density dynamical structure factor $S_{D}({\bf q},{\omega})$ for (a) $h/U=0.19$ (BCS superfluid) and (b) $h/U=0.1978$ (FFLO  superfluid) along $[0,0]\rightarrow [\pi,0]\rightarrow [\pi,\pi]\rightarrow [0,\pi]\rightarrow[0,0]$ in the BZ with $n=1.0$. The white dashed lines exhibit the solution of ${\rm Re}\Sigma=0$ as a function of ${\bf q}$. }
\end{figure}

\begin{figure}[h!]
\includegraphics[scale=0.48]{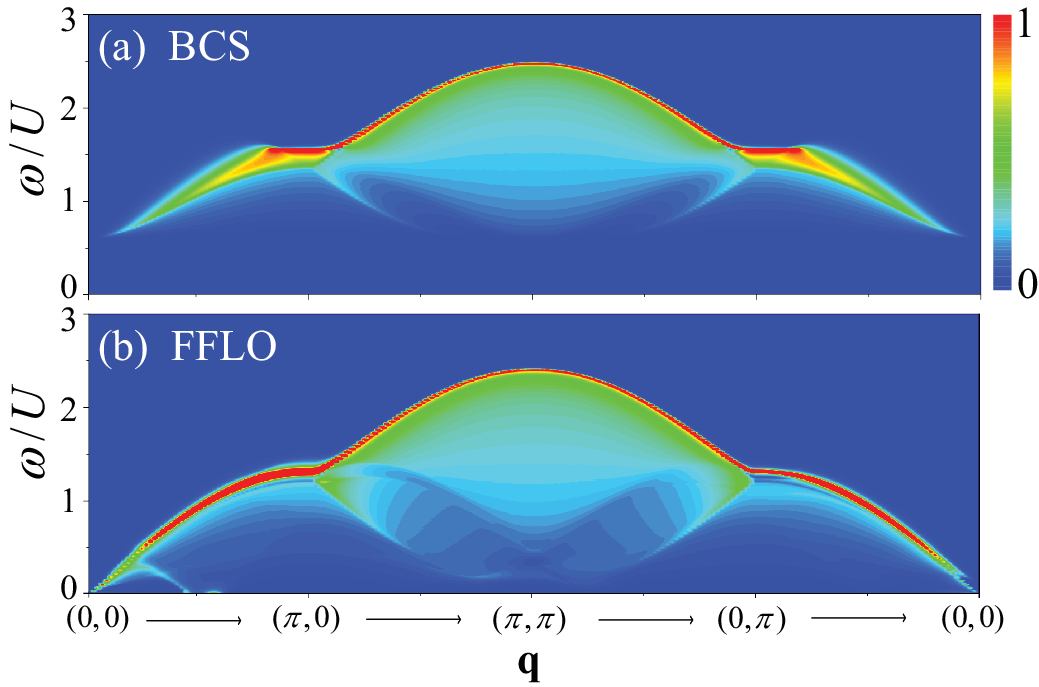}
\caption{Spin dynamical structure factor $S_{S}({\bf q},{\omega})$ (a) $h=0.19$ (BCS superfluid) and (b) $h/U=0.1978$ (FFLO  superfluid). \label{SDSF}}
\end{figure}

 By continuously increasing the Zeeman field strength across the critical value $h_{c}$, the system undergoes a phase transition from a BCS to an FFLO superfluid, resulting in notable changes in both the collective modes and single-particle excitations. In Fig. \ref{DDSF}, we plot the density dynamical structure factor $S_{D}({\bf q},{\omega})$ along the high-symmetry path of the first Brillouin zone (BZ),  $[0,0]\rightarrow [\pi,0]\rightarrow [\pi,\pi]\rightarrow [0,\pi]\rightarrow[0,0]$, for (a) $h/U=0.19$ (BCS superfluid) and (b) $h_{c}/U=0.1978$ (FFLO superfluid), respectively. The corresponding spin dynamical structure factor $S_{S}({\bf q},{\omega})$ is shown in Fig. \ref{SDSF} under the same parameters.  Here, we fix the density at $n=1$ and the hopping strength at $t/U=0.3$.

Obviously, both $S_{D}({\bf q},{\omega})$ and $S_{S}({\bf q},{\omega})$ change dramatically across the the transition from a BCS to an FFLO superfluid, which is mainly attributed to the discontinuous decrease of the pairing gap and the emergence of a finite COM momentum ${\bf Q}$ at the transition point. In the BCS superfluid, $S_{D}({\bf q},{\omega})$ exhibit a sharp collective phonon mode starting from ${\bf q}=[0,0]$, and a roton mode starting from ${\bf q}=[\pi,\pi]$ exhibit a 'V' form. When entering the FFLO superfluid, the collective phonon mode mixes with the gapless single-particle excitations and the roton mode has a 'W' structure. Moreover, at the high energy region, $S_{S}({\bf q},{\omega})$ exhibits a conventional magnetic mode in both the BCS and the FFLO superfluids.

For the BCS state, the single-particle excitations are gapped and the continuum starts at $\omega\ge 2\Delta$. However, for the FFLO superfluid, the quasiparticle energy bands pass through the Fermi energy, giving rise to a Bogoliubov Fermi surface. Another low-energy collective mode associated with the Bogoliubov quasiparticle (bogolon) emerges around ${\bf q}=[0,0]$ in the spin channel. Consequently, the gapless single-particle excitations competes with these collective modes and damps them. Moreover, for an FFLO superfluid, the quasiparticle energy band structure exhibits asymmetry relative to the ${\bf Q}$ direction \cite{Kinnunen2018,Liu2007,Suba2020,Wu2013,Zhou2013}, leading to the anisotropic dynamical excitations compared with the BCS case.

\subsection{Phonon and bogolons}
At small transferred momenta near $\Gamma$ (${\bf q}=[0,0]$), the dynamical structure factors exhibit a sharp narrow peak with a linear dispersion emerging from zero energy, indicating appearance of the collective modes. The phonon mode originates from the spontaneously U(1) phase symmetry breaking of order parameter in the superfluid state  \cite{Nambu1960,Goldstone1962}. In the BCS superfluid, the phonon mode is always well separated from the single-particle continuum. In the FFLO superfluid, however, the phonon mode merges into the single-particle excitation continuum and undergoes spectral broadening through the scattering with the single-particle excitations. In addition to the Cooper pairs, the existence of unpaired, strong spin-polarized atoms near the Bogoliubov Fermi surface give rise to the collective bogolon mode, enhancing the competition between the single-particle excitations and the collective modes.

\begin{figure}[h!]
\includegraphics[scale=0.4]{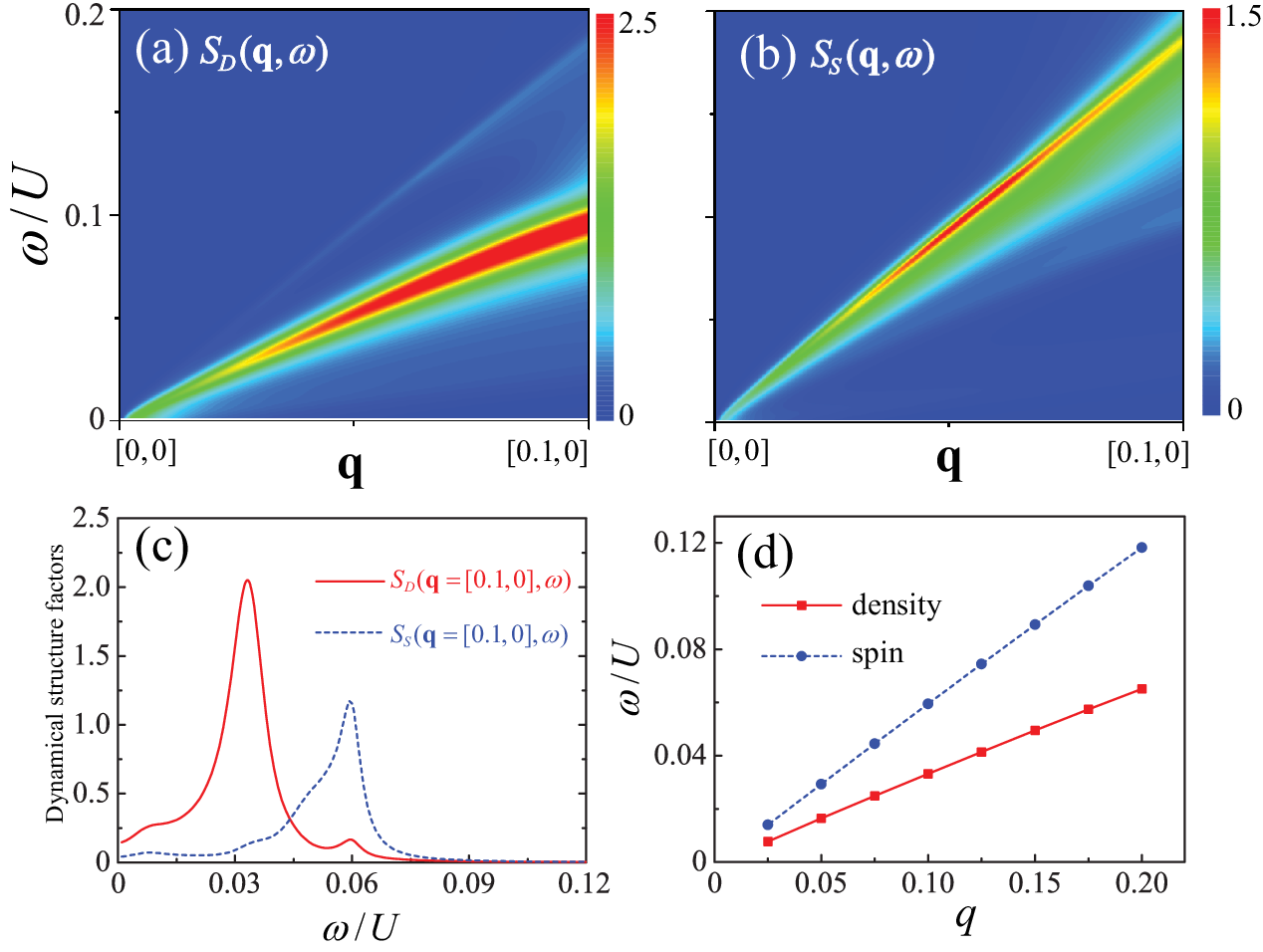}
\caption{(a) $S_{D}({\bf q},{\omega})$ and (b) $S_{S}({\bf q},{\omega})$ of an FFLO superfluid in the low transferred momentum region along $[0,0]\rightarrow [\pi,0]$. (c) $S_{D}({\bf q}=[0.1,0],{\omega})$ (red solid line) and $S_{S}({\bf q}=[0.1,0],{\omega})$ (blue dashed line) as a function of $\omega$; (d) peak positions of two collective modes extracted from $S_{D}({\bf q},{\omega})$ and $S_{S}({\bf q},{\omega})$. Here, the parameters $h/U=0.1978$, $t/U=0.3$, and $n=1$. \label{soundmode}}
\end{figure}

To clearly elucidate these two collective modes in the FFLO superfluid, the zoomed-in $S_{D}({\bf q},{\omega})$ and $S_{S}({\bf q},{\omega})$ in the low-momentum region along $[0,0]\rightarrow[0.1,0]$ are shown in Fig. \ref{soundmode}\textcolor{blue}{(a)} and Fig. \ref{soundmode}\textcolor{blue}{(b)}, respectively.
 Figure \ref{soundmode}\textcolor{blue}{(c)} plot $S_{D}({\bf q}=[0.1,0],{\omega})$ (red solid line) and $S_{S}({\bf q}=[0.1,0],{\omega})$ (blue dashed line) as a function of $\omega$. The extracted peak positions of them are summarized in Fig. \ref{soundmode}\textcolor{blue}{(d)}.

Obviously, Figure \ref{soundmode}\textcolor{blue}{(a)} and Figure \ref{soundmode}\textcolor{blue}{(b)} exhibit two distinct linear collective modes, respectively. A sharp peak in $S_{D}({\bf q}=[0.1,0],{\omega})$ at $\omega/U=0.033$ corresponds to the collective phonon mode. Another peak found in $S_{S}({\bf q}=[0.1,0],{\omega})$ at $\omega/U=0.06$ is a collective bogolon mode. The slope of a collective mode defines the corresponding speed: the sound speed $c_{\rm s}$ in the density channel and the bogolon speed $c_{\rm b}$ in the spin channel. {The unpaired atoms near the Fermi surface are strongly polarized, which explains why the bogolon mode exhibits a stronger signal in the spin dynamical excitations}. The magnitude of these two speeds can be extracted by fitting the position of the $\delta$-like peak of dynamical structure factors at a small transferred momentum, $c_{\rm s/b}=\omega/q$. The bogolon speed $c_{\rm b}/U=0.60$ significantly exceeds that sound speed $c_{\rm s}/U=0.33$. These collective modes can be analyzed through the RPA formula. From Eq. \ref{chi}, the collective modes correspond to the poles of the response function $\chi$, i.e., zeros of the denominator det$[\hat{1}-\chi^{0}({\bf q},{i}\omega_{n})U{M_{I}}]$, which is simplified as $\Sigma=[\chi^{0}_{34}\chi^{0}_{43}-(\chi^{0}_{12})^{2}]$. Following our earlier method  \cite{Zhao2020}, the dispersion of main collective modes is determined by solving the self-consistent equation, ${\rm Re}\Sigma=0$. The solution of ${\rm Re}\Sigma=0$, marked by the white dashed lines, are shown in Fig. \ref{SDSF}\textcolor{blue}{(b)}.

\subsection{Roton mode and COM momentum measurement protocol}
For the BCS superfluid, the roton mode emerges at the momentum point ${\bf q}=[\pi,\pi]$. Its emergence can be attributed to the breaking of a global pseudospin $SU(2)$ symmetry \cite{Zhang1990,Belkhir1994}.  Owing to the particle-hole symmetry at half-filling, the fluctuations of superfluid and CDW become degenerate, leading to the appearance of a gapless roton mode. In contrast, for the FFLO superfluid, the minimum of the roton mode is not localized at a single momentum but forms a ring in the BZ centered at ${\bf q}=[\pi,\pi]$, with a radius $Q=|{\bf Q}|$.

\begin{figure}[h!]
\includegraphics[scale=0.45]{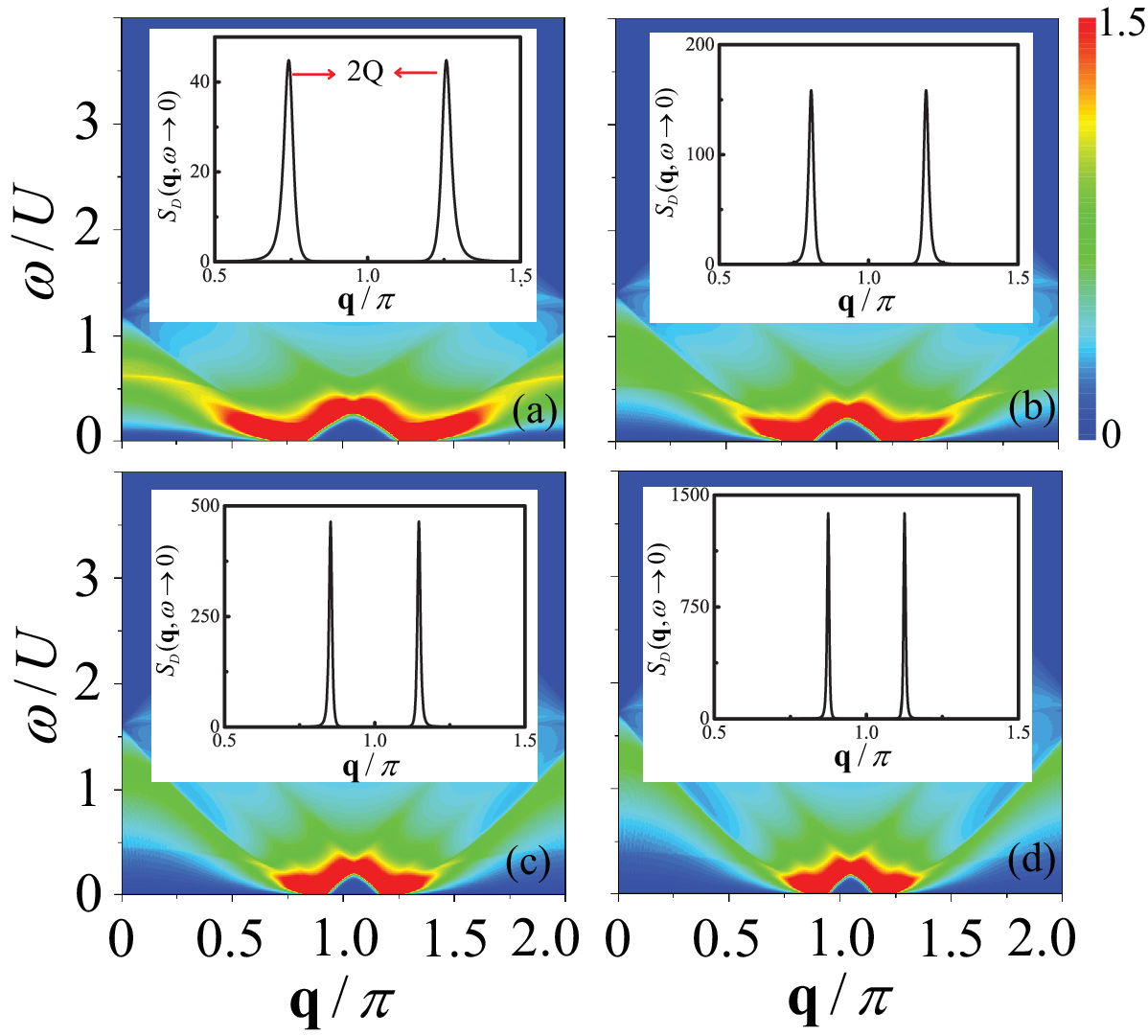}
\caption{ $S_{D}({\bf q},{\omega})$ along the path $[\pi,0]\rightarrow [\pi,\pi]\rightarrow [\pi,2\pi]$ for  (a) $t/U=0.3$, (b) $0.35$, (c) $0.4$, and (d) $0.43$.
\label{rotontj}}
\end{figure}

\begin{figure}[h!]
\includegraphics[scale=0.45]{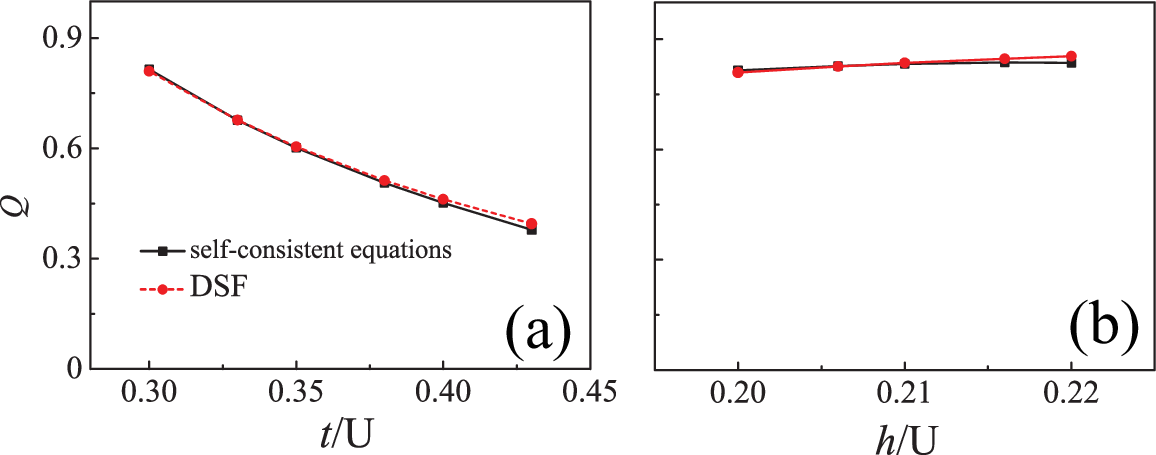}
\caption{ The COM momentum $Q$ as functions of (a) hopping strength $t$ and (b) Zeeman field strength $h$ from the self-consistent calculations (black line) and dynamical structure factor (red line), respectively.
\label{weight}}
\end{figure}

 The ring structure of the roton mode depends sensitively on the interaction strength. In Fig. \ref{rotontj}, we present the contour plots of $S_{D}({\bf q},{\omega})$ along the path $[\pi,0]\rightarrow [\pi,\pi]\rightarrow [\pi,2\pi]$ in the BZ for various hopping strengths.
As $t/U$ increases, the distance by which the roton mode deviates from $[\pi,\pi]$ decreases and becomes approximately equal to the magnitude of COM momentum ${Q}$. To clearly demonstrate the connection between the roton mode and ${Q}$, we extract the displacement of the roton mode from Figs. \ref{rotontj}, and plot $Q$ (red line) as a function of hopping strength $t$ in Fig. \ref{weight}, comparing it with the $Q$ obtained from the self-consistent calculations (black line).
Our results show that the distance of the roton mode deviating from $[\pi,\pi]$ exhibits an almost identical trend for different values of $t$ and $h$.  These results provide an important insight: experimentally, one can directly measure the magnitude of $Q$ by detecting the displacement of the roton mode from $[\pi,\pi]$ in a half-filled lattice system.
\subsection{Angular dependence of dynamical excitations}
\begin{figure}[h!]
\includegraphics[scale=0.45]{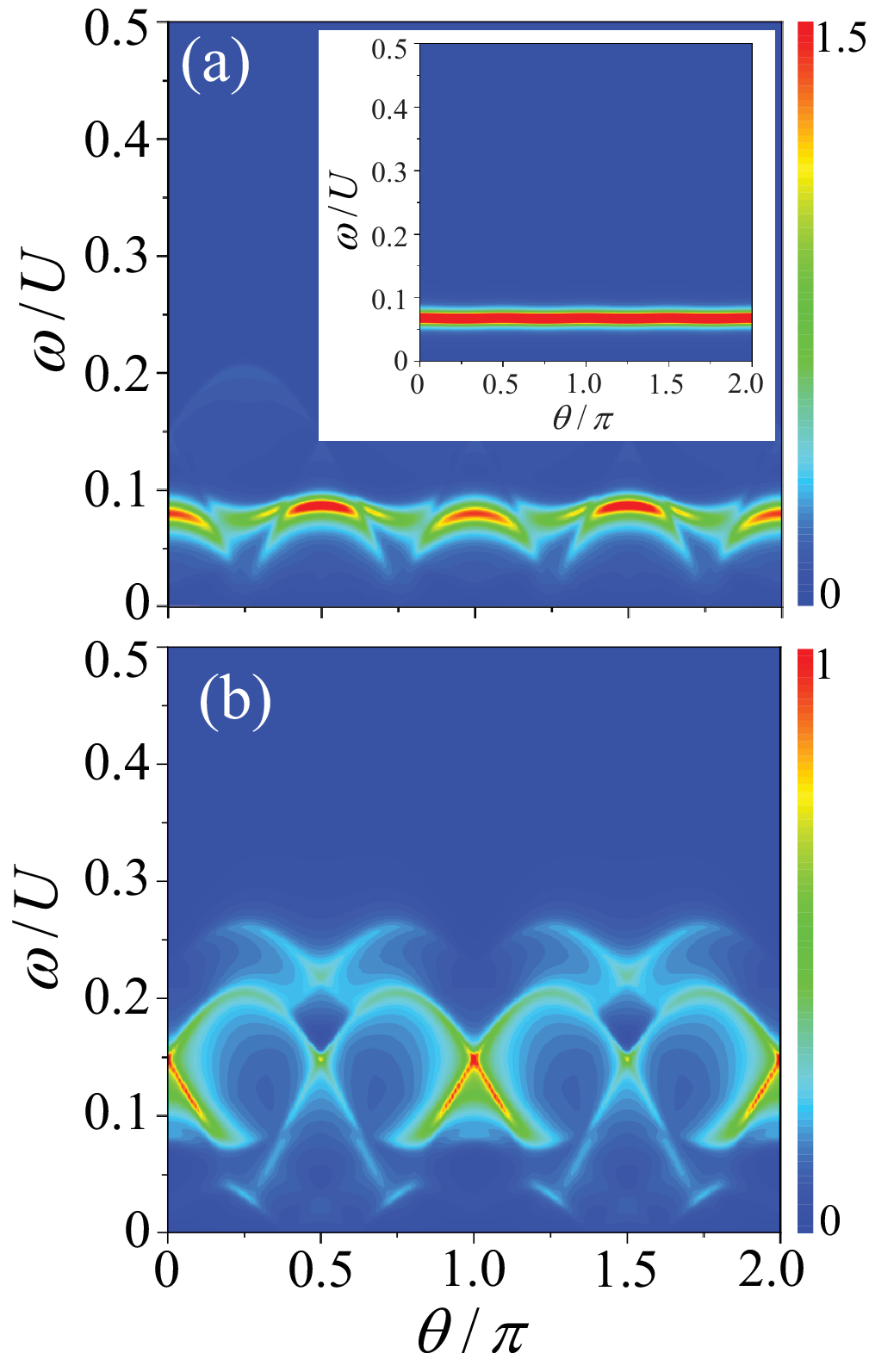}
\caption{(a) $S_{D}({q}=0.08\pi,{\omega})$, (b) $S_{S}({q}=0.08\pi,{\omega})$ in an FFLO superfluid as a function of $\theta$ ranging from $0$ to $2\pi$ with $q=0.08\pi$, $n=1.0$, $h/U=0.1978$. Inset in (a): corresponding to the $S_{D}({q}=0.08\pi,{\omega})$ in the BCS superfluid for $h/U=0.19$.
\label{angular}}
\end{figure}

 To show the anisotropic behavior more clearly, we calculate the angular dependence of the dynamical structure factors at a small transferred momentum. The contour plots of $S_{D}({q}=0.08\pi,{\omega})$ and $S_{S}({q}=0.08\pi,{\omega})$ as a function of $\theta$ are shown in Fig. \ref{angular}, spanning the full angular range.
 In an FFLO superfluid, both $S_{D}({q},\omega)$ and $S_{S}({q},\omega)$ as a function of $\theta$ exhibit an anisotropic behavior, which can be attributed to the finite ${\bf Q}$. Obviously, the strength of the phonon mode evolves with a period of $\pi$ in angle, contrasting sharply with the behavior in a BCS superfluid, where the phonon mode remains nearly constant. Furthermore, the slope of the phonon mode related to the sound speed $c_{\rm s}$ also varies with angle. Notably, unlike a BCS superfluid where the single-particle excitations are gapped by the superfluid gap, the gapless single-particle excitations competes intensely with the phonon mode, and even destroy it at certain angles. Similarly, $S_{S}(\bf{q},\omega)$ displays angular anisotropy, and the bologon mode is also destroyed by the single-particle excitations at certain angles.
\subsection{Single-particle excitations}
\begin{figure}[h!]
\includegraphics[scale=0.48]{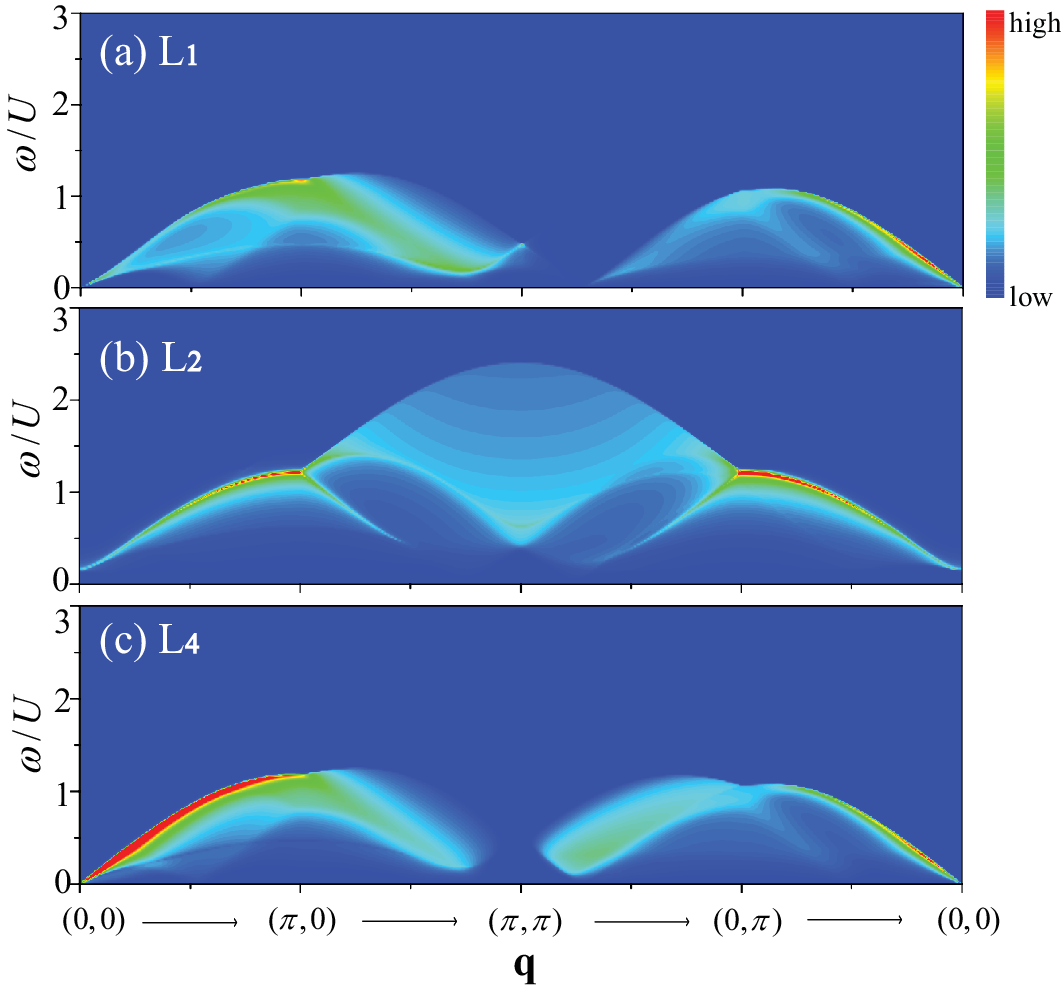}
\caption{The partial color maps of density dynamical structure factor $S_{D}(q,{\omega})$ in Fig (a) to (c) along $[0,0]\rightarrow [\pi,0]\rightarrow [\pi,\pi]\rightarrow [0,\pi]\rightarrow[0,0]$ for $h/U=0.1978$, $n=1.0$.
\label{fig11}}
\end{figure}

 The energy spectra are reconstructed under the Zeeman field, leading to the complex single-particle excitations in an FFLO superfluid, particularly in manifesting the Cooper pair-breaking mechanism. Based on the two quasiparticle spectra $E^{(1)}_{\bf{k}}$, $E^{(2)}_{\bf{k}}$, the dynamical excitations take the following four forms: $\hbar\omega_{kq}=E^{(1)}_{\bf{k+q}}-E^{(1)}_{\bf{k}}, E^{(2)}_{\bf{k+q}}-E^{(2)}_{\bf{k}}, E^{(1)}_{\bf{k+q}}+E^{(2)}_{\bf{k}}, E^{(2)}_{\bf{k+q}}+E^{(1)}_{\bf{k}}$, respectively. These four kinds of excitations are expressed by the kernel functions $L_{1}({\bf k},{\bf q},i\omega_{n})$, $L_{2}({\bf k},{\bf q},i\omega_{n})$, $L_{3}({\bf k},{\bf q},i\omega_{n})$, $L_{4}({\bf k},{\bf q},i\omega_{n})$ in the appendix, where $L_{1}$ and $L_{4}$ are the intra-band excitations while $L_{2}$ and $L_{3}$ are the inter-band excitations.
 Notably, the minima of intra-band excitations remain gapless while the inter-band excitations are gapped.
 To study the contribution of the four excitations, the $S_{D}(q,{\omega})$ at $h=0.2$ is calculated in Fig. \ref{fig11} when only one of kernel functions is considered, (a) $L_{1}({\bf k},{\bf q},i\omega_{n})$, (b) $L_{2}({\bf k},{\bf q},i\omega_{n})$, (c) $L_{4}({\bf k},{\bf q},i\omega_{n})$.
It is shown that the gapless single-particle excitations originate from the intra-band excitations while the inter-band single-particle excitations remain gapped.  The contributions from the inter-band excitations determines the high-energy single-particle spectral features around ${\bf q}=[\pi,\pi]$ region. Moreover, based on our calculations, the contribution from $L_{3}({\bf k},{\bf q},i\omega_{n})$ term is very weak and is therefore omitted here.

\begin{figure}[h!]
\includegraphics[scale=0.45]{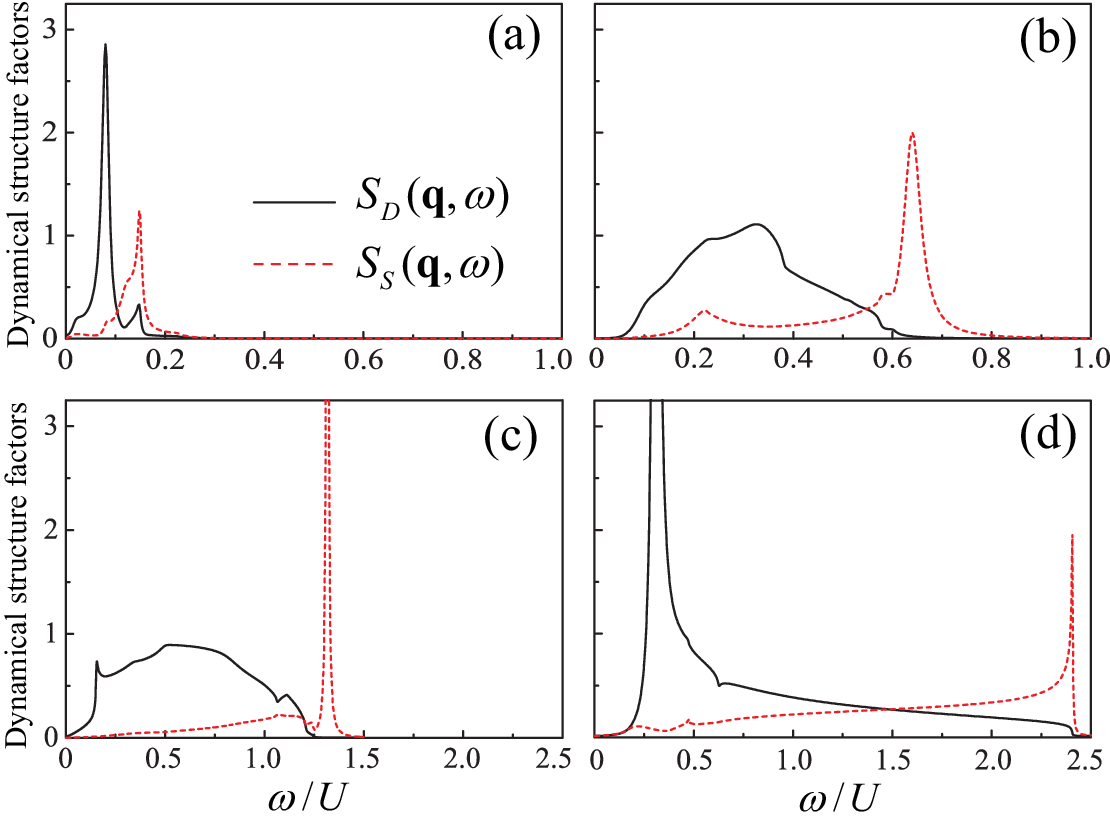}
\caption{ Line cut of $S_{D}({\bf q},{\omega})$ (black solid line) and $S_{S}({\bf q},{\omega})$ (red dashed line) as a function of $\omega$ for (a) $q=[0.08\pi,0]$, (b) $q=[0.32\pi,0]$, (c) $q=[0,\pi]$, and (d) $q=[\pi,\pi]$ with $h/U=0.1978$, $t/U=0.3$.\label{specialpoints}}
\end{figure}

Based on the separate calculation results above, we further calculate the energy dependence of $S_{D}({\bf q},{\omega})$ and $S_{S}({\bf q},{\omega})$ at several typical transferred momenta in the BZ. The results for $S_{D}({\bf q},{\omega})$ (black solid line) and $S_{S}({\bf q},{\omega})$ (red dashed line) are plotted in Fig. \ref{specialpoints}.
At small $q$, a strong, sharp phonon peak appears in the low-energy region, while three kinds of the single-particle excitations at the larger energies are comparatively weak. Owing to the presence of the gapless single-particle excitations, the phonon (bogolon) peak is suppressed and exhibits a finite width, as the phonon (bogolon) mode is pushed into the single-particle excitation region and competes with them. In panel (c), the conventional magnetic mode is separated from the dominant single-particle excitation continuum, producing a sharp magnetic collective peak in $S_{S}({\bf q},{\omega})$. In panel (d), a low-energy sharp peak emerges at $q=[\pi,\pi]$ around $\omega/U=0.31$, reflecting the combined effect of the single-particle excitation and roton-related collective excitations. A higher-energy peak near $\omega/U=2.41$ signals the conventional magnetic mode. Moreover, this excitation gap is doping dependent and will be discussed in the following section.

\section{Doping dependence of dynamical excitations}
\label{DSFDoping}
\begin{figure}[h!]
\includegraphics[scale=0.48]{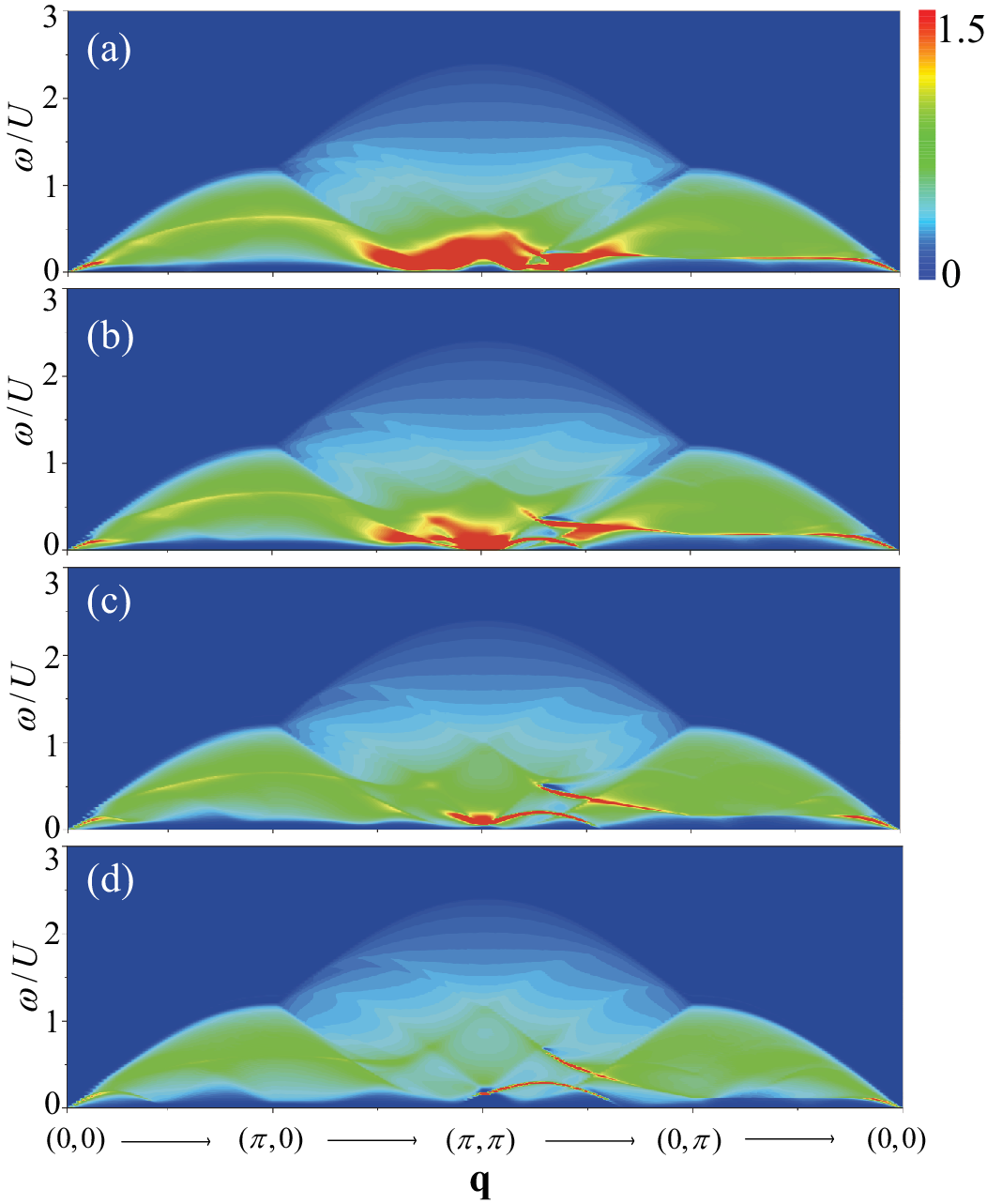}
\caption{Color maps of $S_{D}({\bf q},{\omega})$ for (a) $n=0.9$, (b) $n=0.8$, (c) $n=0.7$, and (d) $n=0.6$ with $t/U=0.3$.
\label{dopingden}}
\end{figure}

\begin{figure}[h!]
\includegraphics[scale=0.48]{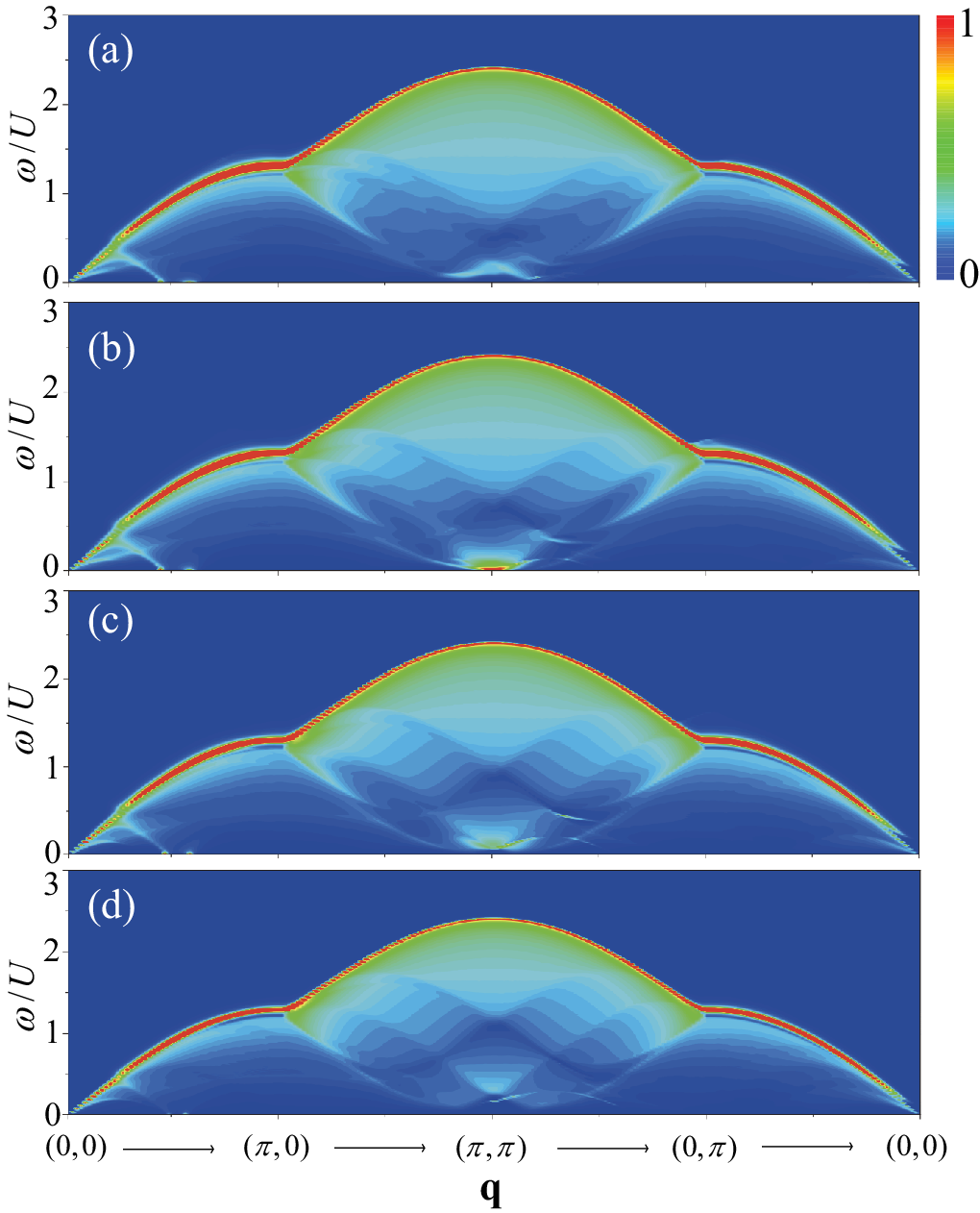}
\caption{Color maps of $S_{S}({\bf q},{\omega})$ for (a) $n=0.9$, (b) $n=0.8$, (c) $n=0.7$, and (d) $n=0.6$ with $t/U=0.3$.
\label{dopingspin}}
\end{figure}

\begin{figure}[h!]
\includegraphics[scale=0.45]{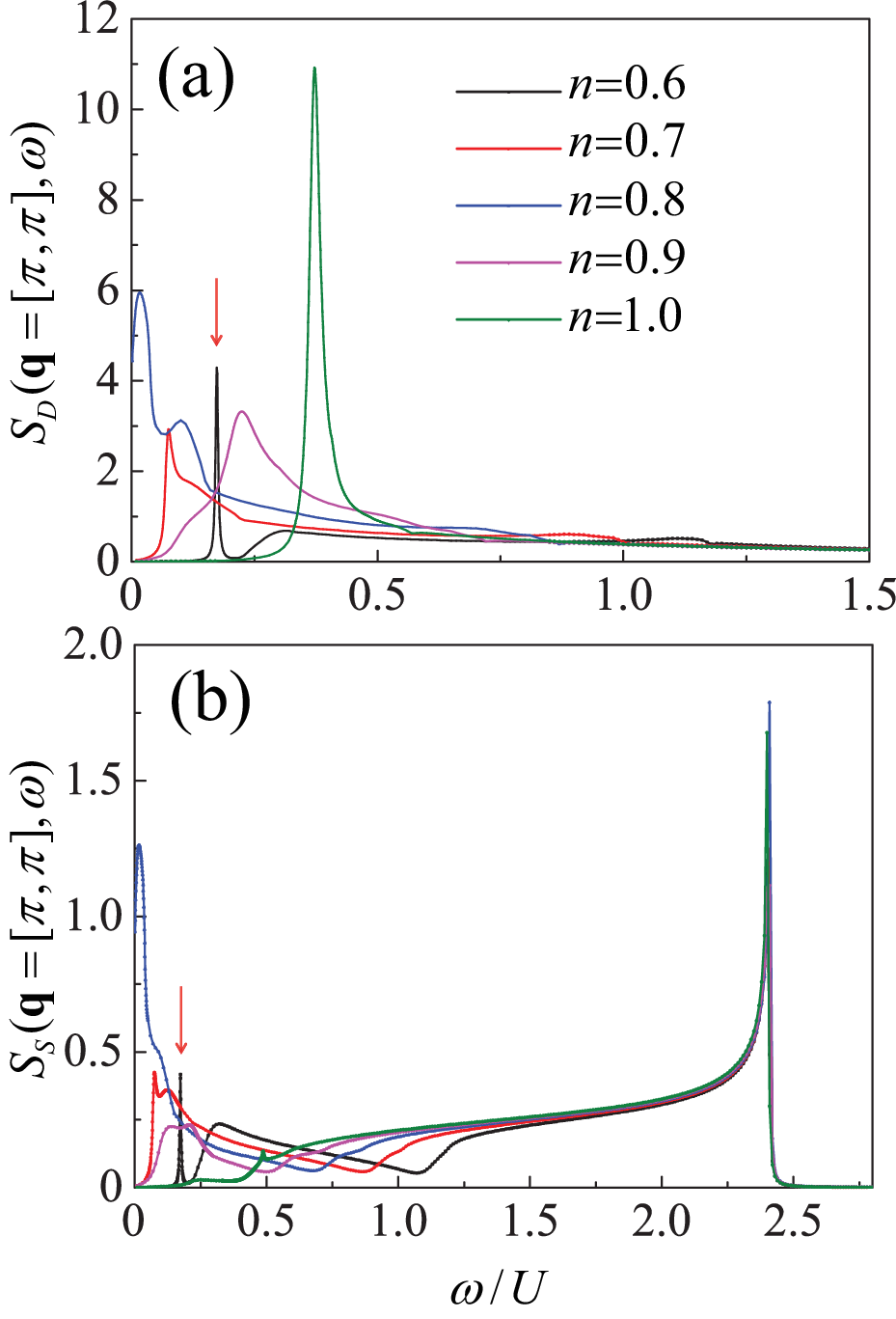}
\caption{(a) $S_{D}({\bf q}=[\pi,\pi],{\omega})$ (b) $S_{S}({\bf q}=[\pi,\pi],{\omega})$ as a function of $n$ ranging at the BCS-FFLO transition point $h_{c}$. \label{pipi-doping}}
\end{figure}

The doping alters the level of Fermi energy. Therefore, we discuss the doping dependent of the dynamical excitations in the FFLO superfluid. In Fig. \ref{dopingden}, we plot $S_{D}({\bf q},{\omega})$ along the high-symmetry directions of the BZ for (a) $n=0.9$, (b) $n=0.8$, (c) $n=0.7$, and (d) $n=0.6$ at the BCS-FFLO transition point $h_{c}$ with $t/U=0.3$. In Fig. \ref{dopingspin}, $S_{S}({\bf q},{\omega})$ as a function of doping are plotted with the same parameters.
To study the dynamical excitations at ${\bf q}=[\pi,\pi]$, we study the doping dependence of dynamical structure factors. As shown in Fig. \ref{pipi-doping}, the dynamical structure factors in the FFLO superfluid are investigated for (a) $S_{D}({\bf q},{\omega})$, and (b) $S_{S}({\bf q},{\omega})$.
Both $S_{D}({\bf q}=[\pi,\pi],{\omega})$ and $S_{S}({\bf q}=[\pi,\pi],{\omega})$ are strongly doping dependent. As $n$ decreases (the doping concentration increases), the excitation gap initially narrows, vanishing around $n=0.8$, and then reopens at lower $n$, exhibiting an open-close-reopen behavior. In particular, at $n=0.6$, the collective mode and the single-particle excitations band are well separated, indicating that the dynamical excitations at ${\bf q}=[\pi,\pi]$ consists of a sharp collective mode (marked by the arrow) and a broad continuum of the single-particle excitations. Furthermore, in Fig. \ref{pipi-doping}\textcolor{blue}{(b)}, the sharp peaks observed around $\omega/U=2.406$ correspond to the conventional magnetic mode, whose positions remain independent of doping concentration.
\section{Discussions}
\label{discussion}
 Following the phase transition from a conventional BCS superfluid to an FFLO superfluid, the dynamical excitations exhibits several novel features, such as the emergence of two distinct collective modes around ${\bf q}=[0,0]$ and a circular roton mode centered at ${\bf q}=[\pi,\pi]$. On the one hand, in addition to the phonon, when the system enters the FFLO superfluid, the energy bands cross the Fermi energy, giving rise to both the gapless single-particle excitations and bogolon mode. This leads to the intense competitions between the single-particle excitations and the collective modes, which shorten or even eliminate the lifetime of the collective modes, making them challenging to observe experimentally. In a conventional BCS superfluid, the Meissner effect ensures an equal population of the spin-up and spin-down atoms, resulting in a non-polarized state. In contrast, in an FFLO state, the applied Zeeman field induces a population imbalance between spin-up and spin-down atoms on the Fermi surface, leading to a net spin polarization and the appearance of the bogolon mode. On the other hand, in an FFLO superfluid, the roton mode forms a ring structure centered at ${\bf q}=[\pi,\pi]$ with a radius equal to COM momentum $Q$, indicating $Q$ directly modifies the dynamical excitations. The finite $Q$ also indicates that the FFLO superfluid exhibits a different symmetry-breaking physics compared to the BCS superfluid. In our previous work on a BCS superfluid at half-filling on an optical lattice (where $Q=0$), the roton mode is found to be gapless and located precisely at $q=[\pi,\pi]$  owing to the particle-hole symmetry. In the FFLO superfluid studied here, however, the particle-symmetry is broken, causing the movement of the roton mode from the $[\pi,\pi]$.
\section{Summary}
\label{summary}
In conclusion, we have systematically discussed the dynamical excitations of an FFLO superfluid state in a 2D spin-polarized Fermi optical lattices system by calculating the dynamical structure factors within the RPA framework. Our analysis reveals, in addition to the phonon mode in the density channel, a distinct bogolon mode is found in the spin channel of the FFLO superfluid. Notably, the evolution of the roton mode provides a novel signature of the finite COM momentum $Q$. We propose that measuring the displacement of this roton mode from $[\pi,\pi]$ point offers a direct experimental protocol to determine $Q$ at the half-filling. The presence of $Q$ also leads to the obvious anisotropic behavior in both single-particle excitations and the collective modes. These characteristic dynamical excitations may provide a potential fingerprint for identifying the FFLO superfluid through the two-photon Bragg spectroscopy. Finally, we have shown that the roton mode and its excitation gap depend strongly on the doping. The roton mode is substantially modified as the doping concentration increases, which restricts the new protocol to measure the COM momentum.
\section{Acknowledgements}
The authors would like to thank Feng Yuan for helpful discussions. This work was supported by the funds from the Research Foundation of Yanshan University under Grant No. 8190448 (S.T.), the National Natural Science Foundation of China under Grant Nos. U23A2073 (P.Z.) and 11547034 (H.Z.).

\section{Appendix}
 The mean-field response function $\chi^0$ of 2D FFLO Fermi superfluid on an optical lattice is numerically calculated, and all 10 independent matrices elements of $\chi^0$ are displayed as,
\begin{eqnarray}\label{a11}
\chi^{0}_{11}&=&\sum_{\bf k}U_{\bf k+q}^2[U_{\bf k}^2L_1({\bf k},{\bf q},i\omega_n)+V_{\bf k}^2L_2({\bf k},{\bf q},i\omega_n)] \nonumber \\
&-&\sum_{\bf k}V_{\bf k+q}^2[{U_{\bf k}^2}L_3({\bf k},{\bf q},i\omega_n)-{V_{\bf k}^2}L_4({\bf k},{\bf q},i\omega_n)],\nonumber
\end{eqnarray}
\begin{eqnarray}\label{a12}
\chi^{0}_{12}&=&-\sum_{\bf k}\frac{\Delta_0^2}{4E_{\bf k}E_{\bf k+q}}[L_5({\bf k},{\bf q},i\omega_n)-L_6({\bf k},{\bf q},i\omega_n)] \nonumber \\
&-&\sum_{\bf k}\frac{\Delta_0^2}{4E_{\bf k}E_{\bf k+q}}[L_7({\bf k},{\bf q},i\omega_n)+L_8({\bf k},{\bf q},i\omega_n)],\nonumber
\end{eqnarray}
\begin{eqnarray}\label{a13}
\chi^{0}_{13}&=&\sum_{\bf k}\frac{\Delta_0}{2E_{\bf k+q}}[{U_{\bf k}^2}L_1({\bf k},{\bf q},i\omega_n)+{V_{\bf k}^2}L_2({\bf k},{\bf q},i\omega_n)] \nonumber \\
&+&\sum_{\bf k}\frac{\Delta_0}{2E_{\bf k+q}}[{U_{\bf k}^2}L_3({\bf k},{\bf q},i\omega_n)-{V_{\bf k}^2}L_4({\bf k},{\bf q},i\omega_n)], \nonumber
\end{eqnarray}
\begin{eqnarray}\label{a14}
\chi^{0}_{14}&=&\sum_{\bf k}\frac{\Delta_0}{2E_{\bf k}}{U_{\bf k+q}^2}[L_1({\bf k},{\bf q},i\omega_n)-L_2({\bf k},{\bf q},i\omega_n)] \nonumber \\
&-&\sum_{\bf k}\frac{\Delta_0}{2E_{\bf k}}{V_{\bf k+q}^2}[L_3({\bf k},{\bf q},i\omega_n)+L_4({\bf k},{\bf q},i\omega_n], \nonumber
\end{eqnarray}
\begin{eqnarray}\label{a22}
\chi^{0}_{22}&=&\sum_{\bf k}V_{\bf k}^2[{V_{\bf k+q}^2}L_5({\bf k},{\bf q},i\omega_n)+{U_{\bf k+q}^2}L_6({\bf k},{\bf q},i\omega_n)] \nonumber \\
&-&\sum_{\bf k}U_{\bf k}^2[{V_{\bf k+q}^2}L_7({\bf k},{\bf q},i\omega_n)-{U_{\bf k+q}^2}L_8({\bf k},{\bf q},i\omega_n)], \nonumber
\end{eqnarray}
\begin{eqnarray}\label{a23}
\chi^{0}_{23}&=&-\sum_{\bf k}\frac{\Delta_0}{2E_{\bf k+q}}{V_{\bf k}^2}[L_5({\bf k},{\bf q},i\omega_n)-L_6({\bf k},{\bf q},i\omega_n)] \nonumber \\
&+&\sum_{\bf k}\frac{\Delta_0}{2E_{\bf k+q}}{U_{\bf k}^2}[L_7({\bf k},{\bf q},i\omega_n)+L_8({\bf k},{\bf q},i\omega_n)], \nonumber
\end{eqnarray}
\begin{eqnarray}\label{a24}
\chi^{0}_{24}&=&-\sum_{\bf k}\frac{\Delta_0}{2E_{\bf k}}[{V_{\bf k+q}^2}L_5({\bf k},{\bf q},i\omega_n)+{U_{\bf k+q}^2}L_6({\bf k},{\bf q},i\omega_n)] \nonumber \\
&-&\sum_{\bf k}\frac{\Delta_0}{2E_{\bf k}}[{V_{\bf k+q}^2}L_7({\bf k},{\bf q},i\omega_n)-{U_{\bf k+q}^2}L_8({\bf k},{\bf q},i\omega_n)], \nonumber
\end{eqnarray}
\begin{eqnarray}\label{a34}
\chi^{0}_{34}&=&\sum_{\bf k}U_{\bf k+q}^2[{V_{\bf k}^2}L_1({\bf k},{\bf q},i\omega_n)+{U_{\bf k}^2}L_2({\bf k},{\bf q},i\omega_n)] \nonumber \\
&-&\sum_{\bf k}V_{\bf k+q}^2[{V_{\bf k}^2}L_3({\bf k},{\bf q},i\omega_n)-{U_{\bf k}^2}L_4({\bf k},{\bf q},i\omega_n)], \nonumber
\end{eqnarray}
\begin{eqnarray}\label{a43}
\chi^{0}_{43}&=&\sum_{\bf k}V_{\bf k}^2[{U_{\bf k+q}^2}L_5({\bf k},{\bf q},i\omega_n)+{V_{\bf k+q}^2}L_6({\bf k},{\bf q},i\omega_n)] \nonumber \\
&-&\sum_{\bf k}U_{\bf k}^2[{U_{\bf k+q}^2}L_7({\bf k},{\bf q},i\omega_n)-{V_{\bf k+q}^2}L_8({\bf k},{\bf q},i\omega_n)].\nonumber \\
\end{eqnarray}
The corresponding functions  $L_1({\bf k},{\bf q},i\omega_n)$, $L_2({\bf k},{\bf q},i\omega_n)$, $L_3({\bf k},{\bf q},i\omega_n)$, $L_4({\bf k},{\bf q},i\omega_n)$, $L_5({\bf k},{\bf q},i\omega_n)$, $L_6({\bf k},{\bf q},i\omega_n)$, $L_7({\bf k},{\bf q},i\omega_n)$, and $L_8({\bf k},{\bf q},i\omega_n)$ are shown as
\begin{eqnarray}\label{kenal1}
L_1({\bf k},{\bf q},i\omega_n) &=& \frac{n_F(E^{(1)}_{\bf k})-n_F(E^{(1)}_{\bf k+q})}{i\omega_n+E^{(1)}_{\bf k}-E^{(1)}_{\bf k+q}} \nonumber
\end{eqnarray}
\begin{eqnarray}\label{kenal1-1}
L_2({\bf k},{\bf q},i\omega_n) &=& \frac{1-n_F(E^{(2)}_{\bf k})-n_F(E^{(1)}_{\bf k+q})}{i\omega_n-E^{(2)}_{\bf k}-E^{(1)}_{\bf k+q}} \nonumber
\end{eqnarray}
\begin{eqnarray}\label{kenal2}
L_3({\bf k},{\bf q},i\omega_n) &=& \frac{1-n_F(E^{(1)}_{\bf k})-n_F(E^{(2)}_{\bf k+q})}{i\omega_n+E^{(1)}_{\bf k}+E^{(2)}_{\bf k+q}} \nonumber
\end{eqnarray}
\begin{eqnarray}\label{kenal2-1}
L_4({\bf k},{\bf q},i\omega_n) &=& \frac{n_F(E^{(2)}_{\bf k+q})-n_F(E^{(2)}_{\bf k})}{i\omega_n-E^{(2)}_{\bf k}+E^{(2)}_{\bf k+q}} \nonumber
\end{eqnarray}
\begin{eqnarray}\label{kenal3-1}
L_5({\bf k},{\bf q},i\omega_n) &=& \frac{n_F(E^{(1)}_{\bf Q-k-q})-n_F(E^{(1)}_{\bf Q-k})}{i\omega_n-E^{(1)}_{\bf Q-k}+E^{(1)}_{\bf Q-k-q}} \nonumber
\end{eqnarray}
\begin{eqnarray}\label{kenal3}
L_6({\bf k},{\bf q},i\omega_n) &=& \frac{1-n_F(E^{(2)}_{\bf Q-k-q})-n_F(E^{(1)}_{\bf Q-k})}{i\omega_n-E^{(1)}_{\bf Q-k}-E^{(2)}_{\bf Q-k-q}} \nonumber
\end{eqnarray}
\begin{eqnarray}\label{kenal4-1}
L_7({\bf k},{\bf q},i\omega_n) &=& \frac{1-n_F(E^{(1)}_{\bf Q-k-q})-n_F(E^{(2)}_{\bf Q-k})}{i\omega_n+E^{(2)}_{\bf Q-k}+E^{(1)}_{\bf Q-k-q}} \nonumber
\end{eqnarray}
\begin{eqnarray}\label{kenal4}
L_8({\bf k},{\bf q},i\omega_n) &=& \frac{n_F(E^{(2)}_{\bf Q-k})-n_F(E^{(2)}_{\bf Q-k-q})}{i\omega_n+E^{(2)}_{\bf Q-k}-E^{(2)}_{\bf Q-k-q}},
\end{eqnarray}
where the function $n_F(x)$ is the Fermi distribution.

\end{document}